\documentclass[prl,twocolumn]{revtex4}
\usepackage{graphicx}
\usepackage{picture}
\usepackage{xcolor}
\usepackage{color}
\usepackage{tikz}

\begin{document}

\title{Using an artificial electric field to create the analog of the red spot of Jupiter in light-heavy  Fermi-Fermi mixtures of ultracold atoms}

\author{H. F. Fotso$^{1}$, J. Vicente$^{2}$, and J. K. Freericks$^{1}$}
\affiliation{$^{1}$Department of Physics, Georgetown University, 37th and O Sts. NW, Washington, DC 20057, USA\\
$^{2}$ Montgomery Blair High School, 51 University Blvd East, Silver Spring, MD 20901, USA}

%$^{2}$Magnet Program, Montgomery Blair High School, 51 University Blvd East, Silver Spring, MD 20901, USA}

\begin{abstract}
Time-of-flight images are a common tool in ultracold atomic experiments, employed to determine the quasimomentum 
distribution of the interacting particles. If one introduces a constant artificial electric field, then the 
quasimomentum distribution evolves in time as Bloch oscillations are generated in the system and then damped showing a complex series of patterns. 
Surprisingly, in different mass Fermi-Fermi mixtures, these patterns can survive for long times, and resemble the
stability of the red spot of Jupiter in classical nonlinear hydrodynamics. In this work, we illustrate the rich
phenomena that can be seen in these driven quantum systems.
\end{abstract}

\maketitle

%==========BODY OF PAPER =========================================

%

\begin{figure}[htbp]
    \begin{center}
      \setlength{\unitlength}{1cm}
      \begin{picture}(8.0,15.0)(0,0)

        \put(0.0, 7.5){\includegraphics*[width=8.0cm, height=7.5cm]{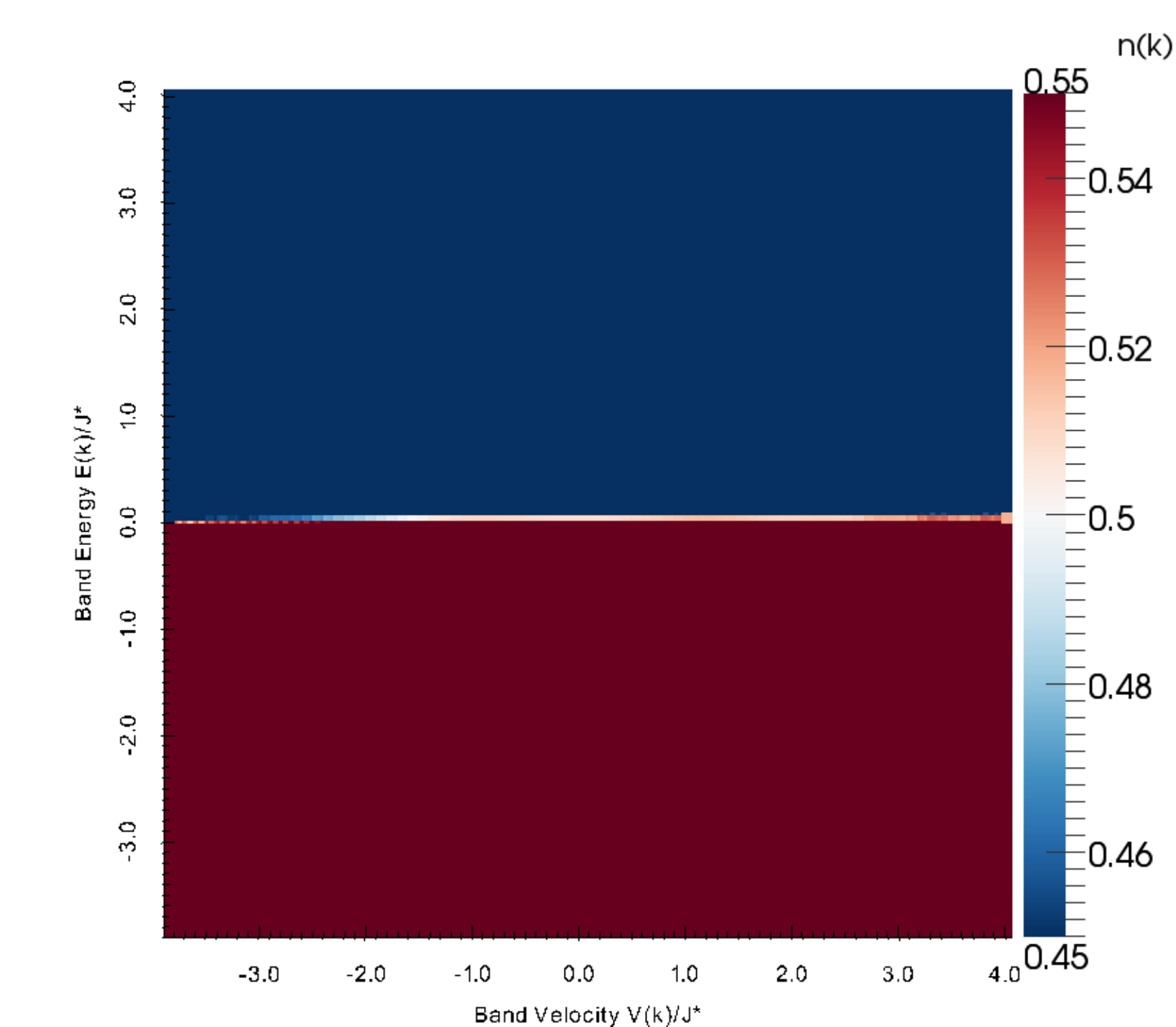}}  
        \put(0.0, 0.0){\includegraphics*[width=7.5cm, height=7.0cm]{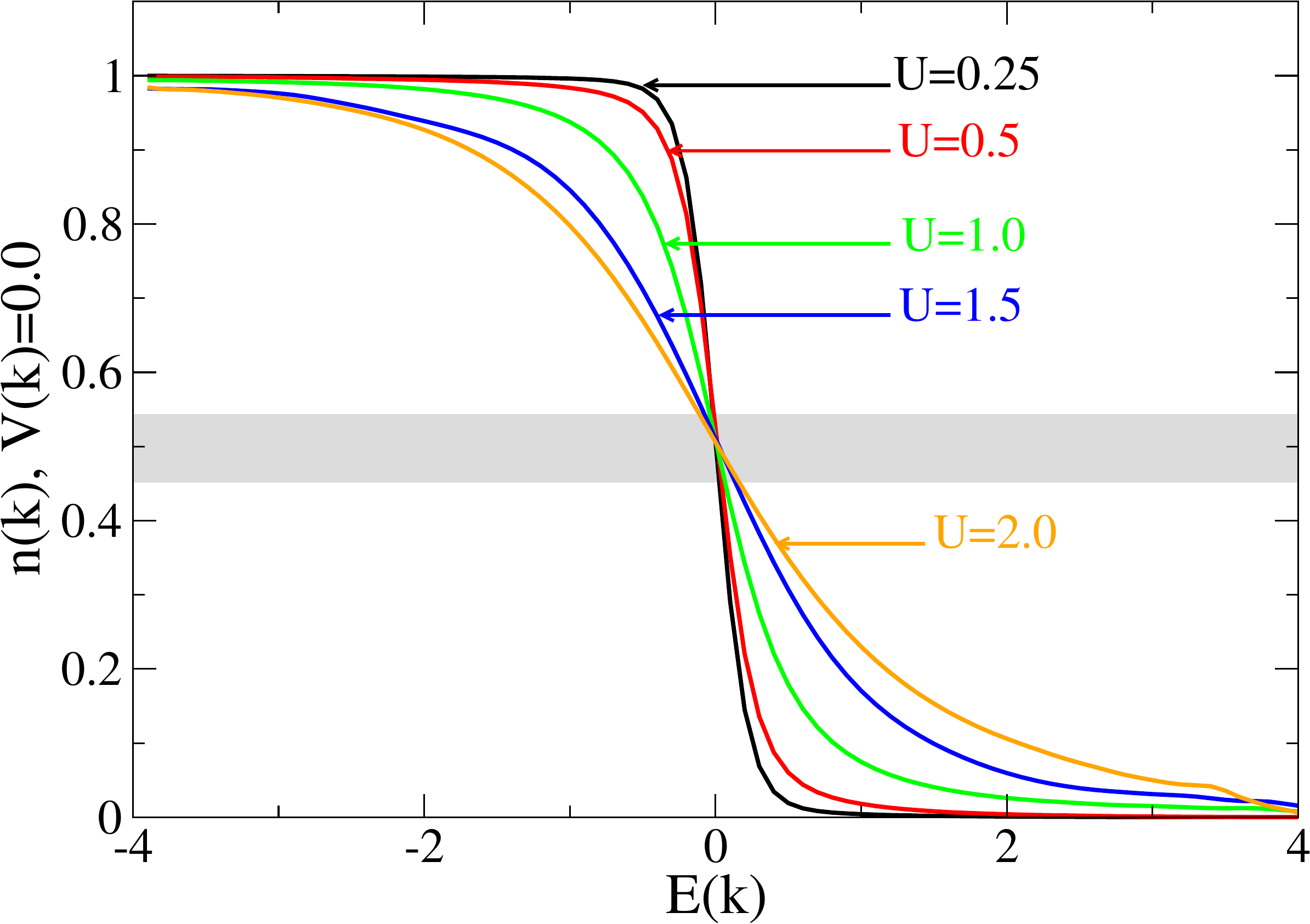}} 

	\put(1.20, 14.50){$t < t_0$}

      \end{picture}
\caption{(Color online) 
False-color image of the initial equilibrium Wigner distribution function at $T=0.1$.  
The upper panel plots $n_{\mathbf k}$ as a function of $E(k)$ and $V(k)$ for $U=0.25$.
The lower panel shows $n_{\mathbf k}$ as a function of $E(k)$ for $V(k)=0.0$ and different values of $U$; the shaded area represents the range 
$[0.45, 0.55]$ over which $n_{\mathbf k}$ is plotted in the upper panel. The deviation from the Fermi-Dirac distribution for $\beta=0.1$ comes from the many-body effects of the interactions between the two types of atoms.
} 
\label{fig:Equilibrium}
      \end{center}
\end{figure}

\paragraph*{Introduction.}
\label{sec:formalism}

The system we study is a light-heavy Fermi-Fermi mixture of repulsively interacting atoms placed on an optical lattice which is 
described in equilibrium by the Falicov-Kimball model~\cite{FalicovKimball,AtesZiegler}:
\begin{eqnarray}
{\mathcal H}_{eq}  = & - & \frac{1}{2\sqrt{d}} \sum_{<ij>} J^*_{ij} \left( c^{\dagger}_{i}c_{j} + c^{\dagger}_{j}c_{i} \right) \nonumber \\  
& - & \mu \sum_{i}  c^{\dagger}_{i}c_{i}  +  U\sum_{i}w_{i}c^{\dagger}_{i}c_{i}
\end{eqnarray}
where $c^{\dagger}_{i}(c_{i})$ is the creation (annihilation) operator for a light fermionic atom at site $i$, 
$w_{i}=0, 1$ is the occupation number operator of a heavy fermionic atom at site $i$, 
$U$ is the s-wave interspecies interaction for doubly occupied lattice sites, $J^*_{ij}=J^*$ is the rescaled  hopping amplitude between nearest-neighbor 
sites $i$ and $j$ of the light atoms and is used as the energy unit; $\mu$ is the light atom
chemical potential. The system is solved in the infinite-dimensional limit of the hypercubic lattice ($d\rightarrow\infty$, noninteracting density of states is a Gaussian~\cite{DMFT}) with dynamical mean-field theory (DMFT)~\cite{DMFT_FK}, which is often a good approximation to three-dimensional systems, at least if the temperature is not too low; we ignore the trap in this work. The system is initially in equilibrium at a temperature $T=0.1$ (in units of $J^*$), then a constant (artificial) electric 
field ${\mathcal E}$ along the diagonal of the hypercubic lattice with ${\mathcal E} = {\mathcal E}(1,1,1,...)$ is switched on at time $t=t_0$ and kept constant for subsequent times.
The uniform electric field ${\mathcal E}(t)$ can be described by a vector potential only gauge with nonzero ${\mathbf A}(t)$ and
${\mathcal E}(t) = -\partial {\mathbf A}(t)/\partial t$ regardless of how it is created in an experiment. 
This is incorporated in the time-dependent Hamiltonian via the Peierls 
substitution~\cite{Peierls}:
\begin{equation}
 J^*_{ij} \to J^*_{ij} {\mathrm {exp}}\left[-i \int_{{\mathbf R}_i}^{{\mathbf R}_j} {\mathbf A}(t) d{\mathbf r}\right] .
\end{equation}
Note that we choose our units so that $c=e=\hbar=1$. Where $c$ is the speed of light, $e$  is the artificial charge, and
$\hbar$ is Planck's constant.

Each atomic species is chosen to have only one spin state, so we can ignore the intraspecies interaction. We consider the model at half-filling for each atomic species ($\rho_{light}=\rho_{heavy}=0.5$), where it obeys particle-hole symmetry. The problem 
is solved using nonequilibrium DMFT
\cite{noneq,FK_NonEq_DMFT08}. Our solutions are formulated in the Kadanoff-Baym-Keldysh formalism 
\cite{BaymKadanoff62, Keldysh64_65} where observables are related to various two time Green's functions 
among which the retarded Green's function is given by Eq.~(\ref{eq:Gretarded}) and the lesser Green's function is given by 
Eq.~(\ref{eq:Glesser}) as follows:
\begin{equation}
 \label{eq:Gretarded}
 G^R_{\mathbf k}(t, t') = -i \, \theta(t, t') {\mathbf {Tr}} \, {\mathcal {T}_c} \, {\mathbf {e}}^{-\beta {\mathcal H}_{eq}}\{c_{\mathbf k}(t),c^{\dagger}_{\mathbf k}(t')\}_+/{\mathcal Z}_{eq};
\end{equation}
\begin{equation}
 \label{eq:Glesser}
 G^<_{\mathbf k}(t, t') = i\, {\mathbf {Tr}} \, {\mathbf {e}}^{-\beta {\mathcal H}_{eq}}c^{\dagger}_{\mathbf k}(t')c_{\mathbf k}(t)/{\mathcal Z}_{eq}.
\end{equation}
Here $\mathbf k $ is a quasimomentum vector in the Brillouin zone, ${\mathcal T}_c$ is the time ordering operator on the Kadanoff-Baym-Keldysh contour, 
$\beta=1/T$ is the inverse temperature, ${\mathcal Z}_{eq}$ is the equilibrium partition function and $\{\cdot,\cdot\}_+$ denotes the anticommutator. (In the remainder of the article we will refer to quasimomentum as momentum.) The retarded Green's function 
determines the quantum-mechanical states of the system while the lesser Green's function 
determines the way in which atoms are distributed among these states.

In the presence of a constant electric field along the diagonal, the infinite dimensional $\mathbf k$-space can be mapped onto a two 
dimensional space characterized by a band energy $E(k)$ given by Eq.~(\ref{eq:bandEnergy}) and a band velocity $V(k)$ given by 
Eq.~(\ref{eq:bandVelocity}), as follows:
\begin{eqnarray}
\label{eq:bandEnergy}
 E(k)& = & \lim_{d \to \infty} \frac{-J^*}{\sqrt{d}} \sum_{i=1}^d {\cos}(k_i);\\
\label{eq:bandVelocity}
 V(k) & = & \lim_{d \to \infty} \frac{-J^*}{\sqrt{d}} \sum_{i=1}^d {\sin}(k_i)
\end{eqnarray}
where the $k_i$'s denote the coordinate of the momentum vector along the different axes of the hypercubic lattice.

In equilibrium, the noninteracting momentum distribution is given by the Fermi-Dirac distribution function. In nonequilibrium and with interactions, it is given by
the gauge-invariant Wigner distribution $n_{\mathbf k}(t) = -i G^<_{{\mathbf k}+{\mathbf A}(t)}(t, t)$~\cite{jauho} which depends only on two variables and is denoted
$n_{E(k), V(k)}(t)$. We use this quantity to track the occupation of states in 
momentum space as a function of time in different parameter regimes with both $E(k)$ and $V(k)$ between -3.9 and 4.0. 
This corresponds to the observable that would be measured in a time-of-flight experiment.

\begin{figure}[htbp]
\includegraphics*[width=8.0cm]{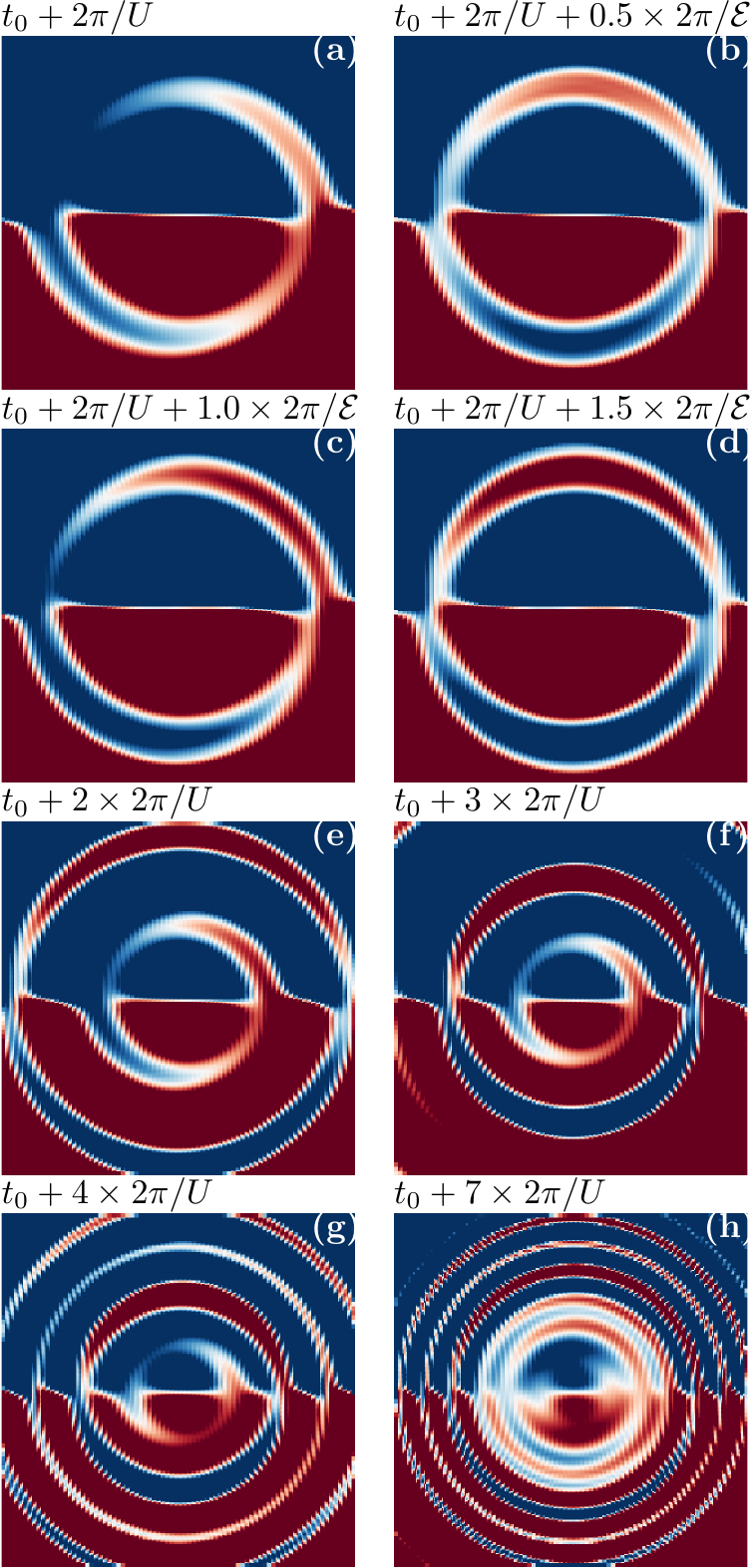}
\caption{(Color online) False color images of the evolution of the gauge-invariant Wigner distribution in momentum space at different times for ${\mathcal E}=2.0$, and $U=0.25$. Each panel shows a snapshot of $n_{\mathbf k}(t)$ at an instant in time 
after  the field is switched on (at $t_0$). Panels (a) through (d) show the evolution for multiples of the timescale $2\pi/{\mathcal E}$ while (e) through
(h) show multiples of the timescale $2\pi/U$.} 
\label{fig:E2p0U0p25}
\end{figure}

\paragraph*{Results.}
Prior to the electric field being switched on, the system is in equilibrium at a temperature $T=0.1$ and the Wigner 
distribution in momentum space is shown in the upper panel of Fig.~\ref{fig:Equilibrium} for $U=0.25$ and is similar for other values of $U$ (but broadened).
Vertical cuts through the data are plotted in the lower panel for $V(k) = 0$ as a function of $E(k)$ for various values of $U$. Note that despite its 
similarity to the Fermi-Dirac distribution of a non-interacting system, this result includes the effects of interactions. The shaded area represents the range $[0.45, 0.55]$ over which $n_{\mathbf k}$ is plotted in the upper panel.
This same range is used for all of the following figures, because it is the typical range over which the structures are seen in the long-time behavior. This 10\% fluctuation in the signal should be measurable in current experiments.

In the case of  a noninteracting system, turning on a static electric field produces Bloch oscillations characterized in frequency space 
by Wannier-Stark ladders \cite{Wannier} and in momentum space by periodic oscillations with a period of $2\pi/{\mathcal E}$ [Fig.~\ref{fig:Equilibrium}(a)]. 
These Bloch oscillations can be measured in the current or other observables \cite{noneq,MeinertNagerl,FK_NonEq_DMFT08, Kolovsky}. 
The time-evolution of the gauge-invariant Wigner distribution in momentum space is simply the rotation of the equilibrium configuration 
around the origin like a clock hand with an unchanged Fermi surface shape.

With the interspecies interaction turned on, the Bloch oscillations are gradually damped and decay towards zero. The Joule heating resulting
from the interaction increases the energy of the system at a rate given by ${\mathcal J}(t) \cdot {\mathcal E}$, with
${\mathcal J}(t)$ being the light atom current~\cite{poles}. This current eventually decays to zero as the isolated system either thermalizes to an infinite
temperature or gets stuck in a nonthermal nonequilibrium steady state. When thermalization occurs, all states are equally occupied 
and we expect $\lim_{t \to \infty} n_{E,V} = 0.5 $ for all $(E, V)$ points \cite{thermalization}. Regardless of the thermalization scenario, 
the long-time limit is approached with the formation of specific patterns depending on the values of the electric field and the 
interspecies  interaction. Two timescales characterize this development. One is related to the Bloch oscillations,
$T_{Bloch}=2\pi/{\mathcal E}$ and the other is related to the collapse and revival of the Bloch oscillations for large fields and small 
interactions, $T_{Beat}=2\pi/U$ \cite{MGreinerBlochNature, MeinertNagerl, BuchleitnerKolovsky}.
We focus here on the large-field case ($\mathcal{E}=2$), which has rich behavior that should be detectable in experiments.

\begin{figure}[htbp]
\includegraphics*[width=8.0cm]{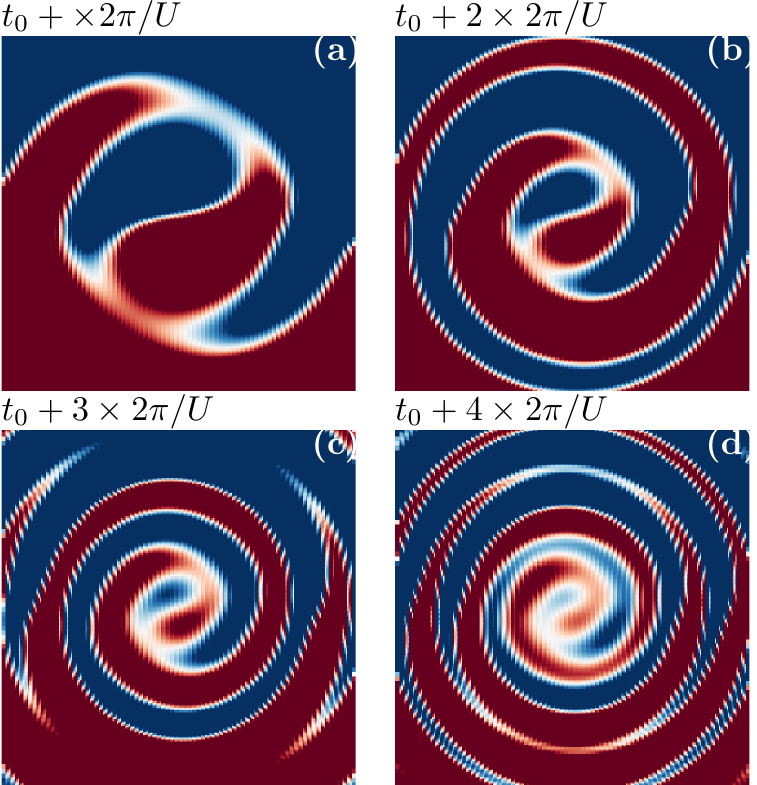}
\caption{(Color online) False color images of the evolution of the gauge-invariant Wigner distribution in momentum space with time for 
${\mathcal E}$=2.0 and  $U=1.0$. Each panel shows a snapshot of $n_{\mathbf k}(t)$ at the corresponding instant in time.
} 
\label{fig:E2p0U1p0}
\end{figure}

Figure~\ref{fig:E2p0U0p25}  shows different stages of the time evolution of the gauge-invariant Wigner distribution in the $(E, V)$-plane
for ${\mathcal E}=2.0$, and $U=0.25$. Panels (a) through (d) probe the timescale $T_{Bloch}$ while (e) through (h) probe the timescale
$T_{Beat}$. This evolution is characterized by the formation of concentric ring shaped turbulences with a region of high occupation 
and a region of low occupation  that are spawned at the origin, $(E, V) = (0,0)$, and then move outward while at the same time 
oscillating around this origin (similar to a pebble being dropped into a pond). These turbulences are formed on a time scale of $T_{Beat}$ reminiscent of the beats that are observed 
in the current as a  function of time for large fields and small interaction values as previously illustrated in 
Ref.~\onlinecite{FK_NonEq_DMFT08}. While they are moving away from the center, these turbulences also subtly rotate around the origin with
a time scale of $T_{Bloch}$. Each new $T_{Beat}$ time interval sees the formation of a new ring at the origin; they eventually pack closer and closer together at long times,
making the region more homogeneous.  Movies of the evolutions are shown with the supplemental material, where it is easier to see the rotations with period $T_{Bloch}$.

In Fig.~\ref{fig:E2p0U1p0}, the evolution is observed for the same value of the electric field, ${\mathcal E}=2.0$ and $U=1.0$. 
In this case, the formation of the rings, their outward motion away from the origin as well as their oscillation around the center  
occur on similar time scales. As a result the rings are no longer separated as in the case of smaller interactions. Instead we see 
the formation of a spiral whose length grows with time (like a dog catching its tail). The spiral grows in length with the addition of a new layer after each $T_{Beat}$
time step in a way analogous to the case of smaller interactions. 
The central region has a persistent pattern similar to the Yin and Yang symbol in Chinese culture.

\begin{figure}[htbp]
\includegraphics*[width=8.0cm]{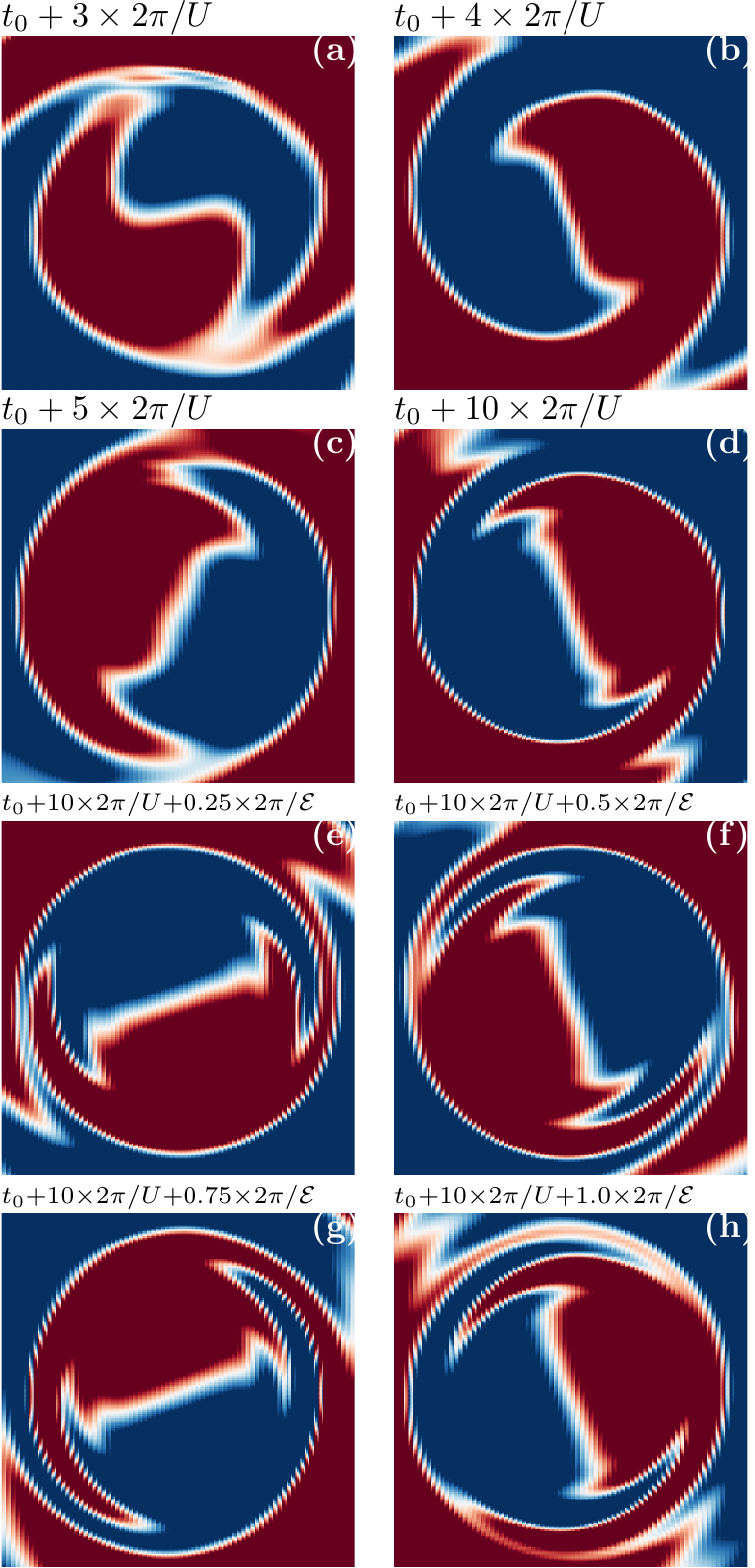}
\caption{(Color online) False color image of the evolution of the gauge-invariant Wigner distribution with time for ${\mathcal E}=2.0$ 
and $U=3.0$. Panels (a) through (d) show the behavior of the system on the timescale $2\pi/U$ while panels (e) through
(h) show the timescale $2\pi/{\mathcal E}$.
} 
\label{fig:E2p0U3p0}
\end{figure}

When the interspecies interaction becomes larger, we see the formation in the middle of the spiral of a feature topologically
analogous to the ``red spot of Jupiter'' with one region of high occupation and one of low occupation (see Fig.~\ref{fig:E2p0U3p0}). 
The formation of rings is initially reduced to that of small sharp edges on two well defined 
regions as shown in Fig.~\ref{fig:E2p0U3p0} (a) through (d). At even longer times, we observe no further changes to this central feature 
and all the disturbances seem to be taking place outside of the simulated region. Throughout this evolution, the whole system rotates 
around the origin with a period of $T_{Bloch}$ as seen in Fig.~\ref{fig:E2p0U3p0} (d) through (h). The size of this ``red spot of Jupiter'' grows with the interaction so that for even larger $U$, one would see a behavior analogous to the rotating Fermi surface 
of the noninteracting system at the Bloch frequency.

The evolution of the gauge-invariant Wigner distribution shows the development of patterns that are robust and persistent up to very long times. 
It is important to note that these patterns are most apparent when one focuses in on a window around the infinite temperature thermal state 
where  $n_{\mathbf k} = 0.5$. Initially, the Wigner distribution is between 0 and 1 and as the system heats up due to Joule heating, 
this interval is gradually reduced so that at long times, the high density region and the low density region are shown with Wigner distributions focused
between 0.45 and 0.55. Note that the diffraction pattern at the boundaries between different regions is due to the rendering algorithm of the 
plotting software \cite{Paraview} and are not physical results.

\paragraph*{Conclusion.}
\label{sec: Conclusion}
We have studied the real-time evolution of the distribution function in momentum space of a field-driven light-heavy Fermi-Fermi mixture of atoms described by the 
Falicov-Kimball model. This evolution is governed by timescales related to Bloch oscillations and beats in the current through the formation of specific patterns of 
the distribution function. The system, initially in equilibrium with a distribution function similar
to a Fermi-Dirac distribution, evolves through these patterns towards a stationary state where all states are equally occupied [$n_{\mathbf k}=0.5$]. For ${\mathcal E}=2.0$ and $U=0.25$, we found that the momentum space distribution function develops concentric ring-shaped 
turbulences around the origin (pebble in the pond). These become a spiral when $U=1.0$ (dog chasing its tail)  and for larger $U$, we see the formation of a feature analogous to the 
``red spot of Jupiter" with a stable region of high occupation and low occupation that rotates around the origin with the Bloch 
period. Current technology in cold atom experiments should be able to see these features.

\paragraph*{Acknowledgments-} JKF and HFF were supported by the National Science Foundation grant No. DMR-1006605.  HFF was additionally supported by the Air Force Office of Scientific Research under MURI grant No. FA9559-09-1-0617 for the latter stages of the work. High performance computer resources utilized resources under a challenge grant from the High Performance Modernization program of the Department of Defense. JKF was also supported by the McDevitt bequest at Georgetown University.

\clearpage
\newpage
\appendix
\begin{widetext}
\begin{center}
\vspace{5mm}
 \textbf{Supplemental Material for:\\
Using an artificial electric field to create the analog of the red spot of Jupiter in light-heavy 
Fermi-Fermi mixtures of ultracold atoms}
\vspace{8mm}
\end{center}
\end{widetext}
\setcounter{equation}{0} 

Here we provide further details about the numerical calculation of the gauge-invariant Wigner distribution function as well as the production 
of the corresponding videos. 

\begin{figure}[htbp]
\includegraphics*[width=8.0cm]{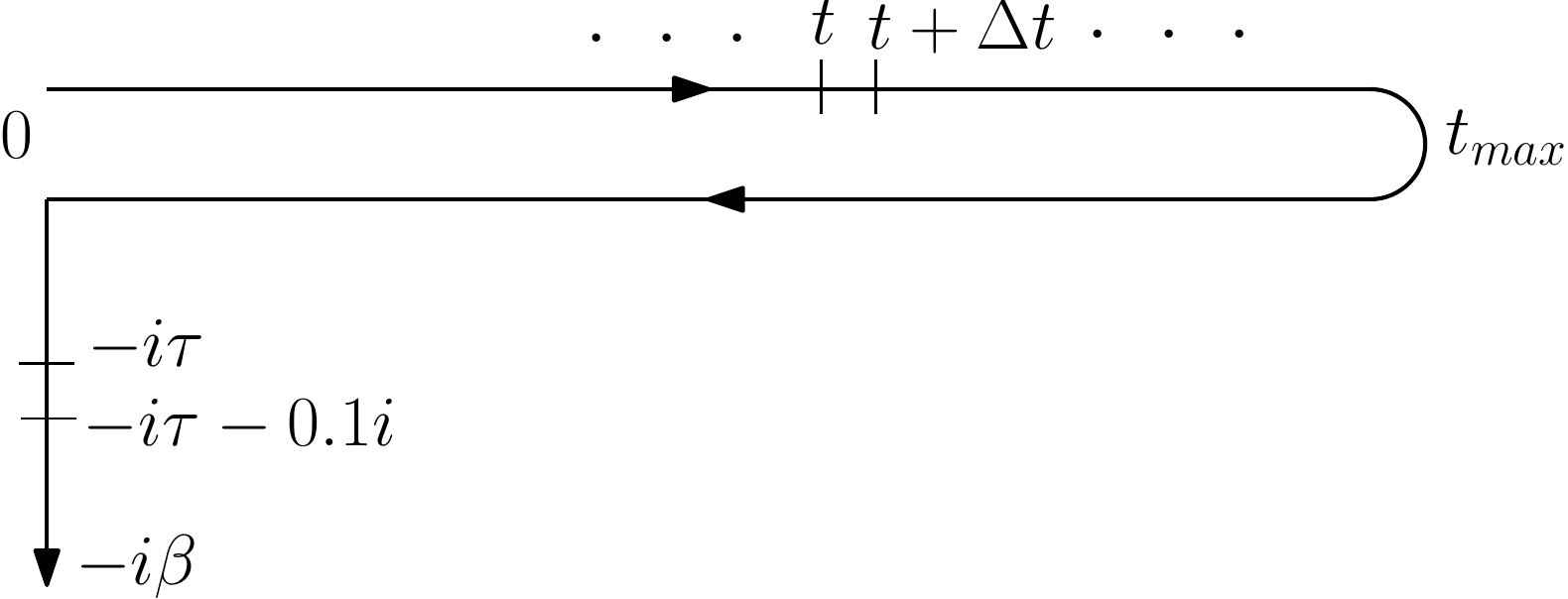}
\caption{Kadanoff-Baym-Keldysh contour for the nonequilibrium calculation of the two-time Green's function.} 
\label{fig:KeldyshContour}
\end{figure}

The nonequilibrium problem is solved on the Kadanoff-Baym-Keldysh contour which is discretized with a spacing $\Delta t$ between consecutive times on the real branch while the imaginary
branch has a spacing of $0.1i$ as shown in Fig.~\ref{fig:KeldyshContour}. The calculation is carried out for different values of $\Delta t$ and then extrapolated using a quadratic 
extrapolation to $\Delta t =0$.

In dynamical mean-field theory, the self-energy keeps its time dependence but has no momentum dependence. This can be 
written as:
\begin{equation}
 \Sigma_{\mathbf k}(t, t') = \Sigma(t, t').
\end{equation}
The DMFT problem is solved for the two-time contour-ordered Green's function defined by 
\begin{equation}
 \label{eq:Gcontour}
 G^c_{\mathbf k}(t, t') = -i\, {\mathbf {Tr}} \, {\mathcal {T}_c} {\mathbf {e}}^{-\beta {\mathcal H}_{eq}}c_{\mathbf k}(t)c^{\dagger}_{\mathbf k}(t')/{\mathcal Z}_{eq}
\end{equation}
with the operators in the Heisenberg representation.
To calculate the local contour-ordered Green's function $G^c_{loc}(t, t') $, a summation over momentum is necessary. The problem is 
first mapped onto the two band energies $E(\mathbf k)$ and $V(\mathbf k)$, so that the
summation is reduced to a double integration. This integral is then calculated with a double Gaussian integration. 
 
The local retarded and the lesser Green's functions, $G^R(t, t')$ and $G^<(t, t')$, are extracted from the local contour ordered Green's 
function. The complete description of this solution is given in Ref.~\cite{FK_NonEq_DMFT08}. In addition, the algorithm also determines the local self-energy. From this, 
the {\textbf k}-dependent retarded and lesser Green's functions, $G^<_{E({\mathbf k}), V({\mathbf k})}(t, t')$ and 
$G^R_{E({\mathbf k}), V({\mathbf k})}(t, t')$ are constructed by using Dyson's equation and the momentum-dependent noninteracting Green's function in the presence of a field. 

Note that these Green's functions are calculated in the vector-potential-only gauge. However, it is the gauge-invariant Wigner distribution~\cite{jauho} that is experimentally measurable. The gauge-invariant Wigner distribution is defined by:
\begin{equation}
n_{E({\mathbf k}), V({\mathbf k})} (t) = -i G^<_{E({\mathbf k} + {\mathbf A}(t)), V({\mathbf k} + {\mathbf A}(t))}(t, t),
\end{equation}
but we actually calculate it from the ratio
\begin{equation}
n_{E({\mathbf k}), V({\mathbf k})}(t) = -i \frac {G^<_{E({\mathbf k} + {\mathbf A}(t)), V({\mathbf k} + {\mathbf A}(t))}(t, t)}{
G^R_{E({\mathbf k} + {\mathbf A}(t)), V({\mathbf k} + {\mathbf A}(t))}(t, t)}
\end{equation}
since the equal-time retarded Green's function is simply the equalt-time anticommutator which is equal to 1. We find this ratio expression to 
converge faster to the $\Delta t \rightarrow0$ result, since the retarded Green's function for a given discretization size often is not precisely equal to 1.

The convergence is generally robust for small interactions and becomes harder for large interactions where a finer time grid is required and where the equilibrium 
result is difficult to reproduce. Moment sum rules extrapolated to $\Delta t=0$ are used to gauge the accuracy of the final calculations.

To produce the movies of the time evolution of the gauge-invariant Wigner distribution, we use the visualization software \textit{paraview}~\cite{Paraview}. The distribution function
$n_{E({\mathbf k}), V({\mathbf k})}(t)$ is represented with a false color plot for each time step as a function of $E(\mathbf k)$ and $V(\mathbf k)$ producing the 
corresponding frame. The frames are then linked together to produce a video. A measure of time is given by the horizontal bar 
at the top of the graph. Initially, it's orange color indicates elapsed time with the system in equilibrium (no electric field). 
This is followed by the coloring in green that indicates elapsed time after the electric field is switched on at $t_0 = 5.0$.

For the animations, we used a region of k-space defined by a disk of radius 4 in the $[E(\mathbf k),V(\mathbf k)]$ plane. This choice is due to the fact that the band energy and
band velocity will be constrained in a similar way for an experiment in finite dimensions and also that it is visually more compatible with the rotations
arising from the electric field.

All the movies presented are for the electric field ${\mathcal E}=2.0$. For $U=0.25$, the system can be tracked up to long times
(here up to $t = 200$) with $\Delta t = 0.1$ and no extrapolation. For $U=1.0$, the time evolution is carried out up to 
$t=40$ and the scaled  result for $\Delta t = 0.0$ is obtained by quadratically extrapolating the data with $\Delta t = 0.066$, $\Delta t = 0.05$ and $\Delta t = 0.04$.
For $U=3.0$, since the equilibrium result is difficult to reproduce, we simply use one frame repeated for times 
prior to the electric field being switched on. For subsequent times, the data is extrapolated to $ \Delta t =0$ using results from $ \Delta t =0.033$, 
$ \Delta t =0.025$ and $ \Delta t =0.02$. This produces a small jump in the data at $t_0=5$

To improve the resolution of the videos, an interpolation is performed for $n(E,V)$ from a grid with spacing $\Delta E = \Delta V = 0.1$ to one with
$\Delta E = \Delta V= 0.025$. This improves the resolution of the images by increasing the pixelation without changing the underlying structure. Note that the rendering algorithim of paraview works by coloring triangulated images, which produces an artificial diffraction effect at some of the boundaries of the images, which is an artifact of this rendering algorithm and not a real effect in the data.

\end{document}